\documentclass{webofc}
\usepackage[varg]{txfonts}   
\begin{document}

\def\planck{{\it Planck}} 
\def\aj{AJ}%
\def\araa{ARA\&A}%
\def\apj{ApJ}%
\def\apjl{ApJ}%
\def\apjs{ApJS}%
\def\aap{Astron. Astrophys.}%
 \def\aapr{A\&A~Rev.}%
\def\aaps{A\&AS}%
\def\mnras{MNRAS}
\def\ssr{SSRv}
\def\nat{Nature}
\def\jcap{JCAP}
\title{Cosmological implications of a modified galaxy cluster pressure profile using the \planck\ tSZ power spectrum}
%
%
          
\author{\firstname{F.} \lastname{Ruppin} \inst{\ref{MIT}}\fnsep\thanks{\email{ruppin@mit.edu}}
\and  \firstname{F.} \lastname{Mayet} \inst{\ref{LPSC}}
\and  \firstname{J.F.} \lastname{Mac\'ias-P\'erez} \inst{\ref{LPSC}}
\and  \firstname{L.} \lastname{Perotto} \inst{\ref{LPSC}}}

\institute{
Kavli Institute for Astrophysics and Space Research, Massachusetts Institute of Technology, Cambridge, MA 02139, USA
\label{MIT}
\and
Univ. Grenoble Alpes, CNRS, Grenoble INP, LPSC-IN2P3, 53, avenue des Martyrs, 38000 Grenoble, France
  \label{LPSC}
}

\abstract{%
The mean pressure profile of the cluster population is a key element in cosmological analyses based on surveys of galaxy clusters observed through the Sunyaev-Zel'dovich (SZ) effect. A variation of both the shape and the amplitude of this profile could explain part of the discrepancy currently observed between the cosmological constraints obtained from the analyses of the CMB primary anisotropies and those from cluster abundance in SZ surveys for a fixed mass bias parameter. We study the cosmological implications of a modification of the mean pressure profile through the analysis of the SZ power spectrum measured by \planck. We define two mean pressure profiles on either side of the one obtained from the observation of nearby clusters by \planck. The parameters of these profiles are chosen to ensure their compatibility with the distributions of pressure and gas mass fraction profiles observed at low redshift. We find significant differences between the cosmological parameters obtained by using these two profiles to fit the \planck\ SZ power spectrum and those found in previous analyses. We conclude that a ${\sim}15\%$ decrease of the amplitude of the mean normalized pressure profile is sufficient to alleviate the discrepancy observed between the constraints of $\sigma_8$ and $\Omega_m$ from the CMB and cluster analyses.
}
\maketitle
\section{Introduction}\label{sec:intro}

The abundance of galaxy clusters as a function of mass and redshift has been shown to be a powerful cosmological probe as it is both sensitive to the statistical properties of the matter distribution at the end of the inflation and to the growth of structures across the whole history of the universe \citep{voi05}. The thermal Sunyaev-Zel'dovich (tSZ) effect is an excellent observable to measure the distribution of galaxy clusters up to very high redshift \citep{boc18}. The calibration of both the mean normalized pressure profile of the cluster population and the scaling relation that relates the tSZ observable and cluster mass is fundamental to establishing accurate cosmological constraints from the analysis of tSZ surveys. The analysis of the tSZ angular power spectrum measured by \planck\ \citep{pla16b} have enabled estimating the amplitude of the linear matter power spectrum at a scale of $8h^{-1}$Mpc, $\sigma_8$, and the total matter density of the universe $\Omega_m$. However, these cosmological constraints are in tension with the ones obtained from the analysis of the power spectrum of the CMB temperature anisotropies \citep{pla16c}.\\
A significant amount of research has been made to show that this tension could be due to a limit in the $\Lambda$CDM model or to a biased estimation of the mass of galaxy clusters. Part of this discrepancy may also be due to a mass and redshift evolution of the mean normalized pressure profile. Indeed, the ones currently used in tSZ cosmological analyses have been estimated using cluster samples at low redshift and high mass \citep{arn10,pla13}. However, deviations from the self-similar hypothesis may lead to significant differences between these profiles and the true mean normalized pressure profile of the cluster population. One of the goals of the NIKA2 \citep{ada18} tSZ large program is to analyze the potential redshift evolution of the mean normalized pressure profile \citep{per18}. In this paper, we study the impact of a modification of the mean normalized pressure profile of the cluster population on the constraints of the $\sigma_8$ and $\Omega_m$ parameters derived from the analysis of the tSZ power spectrum measured by \planck\ \citep{pla16b}.

\section{Cosmology from the tSZ power spectrum}\label{sec:tsz_power_spec}

\subsection{Model of the tSZ power spectrum}\label{subsec:model_tsz_spec}

We model the tSZ power spectrum using only the one-halo component as the two-halo term has been shown to be negligible given the precision of the current tSZ measurements \citep{kom99}. It is given at a multipole $\ell$ by:
\begin{equation}
C_{\ell}^{tSZ} = \int \frac{d^2V}{dzd\Omega} \, dz \int \frac{dn}{dM_{500}} \, \left|\frac{4\pi R_{500}}{\ell_{500}^2}  \frac{\sigma_T}{m_ec^2} \, P_{500} \, I_{\mathbb{P}}(\ell_{500})\right|^2 \, dM_{500} 
\label{eq:ClSZ_cosmo_analysis}
\end{equation}
where $m_{e}$ is the electron mass, $c$ the speed of light, and $\sigma_{\mathrm{T}}$ the Thomson scattering cross section. The characteristic radius $R_{500}$ is the upper integration limit at which the mean cluster density is $500$ times the critical density of the universe. The mass function $dn/dM_{500}$ gives the expected halo number density for a cluster mass $M_{500}$. Its amplitude highly depends on the value of the $\sigma_8$ and $\Omega_m$ parameters. The tSZ power spectrum also slightly depends on the Hubble parameter $H_0$ through the comoving volume element $d^2V/dzd\Omega$. A key element in this model is the $I_{\mathbb{P}}(\ell_{500})$ function giving the normalized two dimensional Fourier transform of the mean pressure profile. It is given under the Limber's approximation by \citep{kom02}:
\begin{equation}
I_{\mathbb{P}} = \int x^2 \frac{\mathrm{sin}(\ell x/\ell_{500})}{\ell x/\ell_{500}} \mathbb{P}(x) \, dx
\label{eq:tabulated_y}
\end{equation}
where $x=r/R_{500}$, $\ell_{500} \equiv D_A/R_{500}$, $D_A$ is the angular diameter distance, and $\mathbb{P}(x)$ is the mean normalized pressure profile of the cluster population. We model this profile using a generalized Navarro-Frenk-White model \citep[gNFW, ][]{nag07}:
\begin{equation}
\mathbb{P}(x) = \frac{P_0}{(c_{500}x)^{\gamma}[1+(c_{500}x)^{\alpha}]^{(\beta - \gamma)/\alpha}}
\label{eq:cosmo_normalized_P}
\end{equation}
This profile is scaled by the $P_{500}$ parameter in Eq. (\ref{eq:ClSZ_cosmo_analysis}) to account for the mass dependence of the cluster pressure content. It has been shown by Bolliet \emph{et al.} \citep{bol18} that the amplitude of the power spectrum of the tSZ effect scales with the combined parameter
\begin{equation}
F = \sigma_8 \, (\Omega_m/B)^{0.40} \, h^{-0.21}
\label{eq:defintion_F}
\end{equation}
where $h = H_0 / [100~\mathrm{km/s/Mpc}]$ and $B = 1/(1-b)$ is the hydrostatic bias that links the true mass $M_{500}$ to the one given under the assumption of hydrostatic equilibrium $M_{500}^{\mathrm{HSE}} = (1-b) M_{500}$. In the following, we aim at quantifying the implications of a modification of the profile given in Eq. (\ref{eq:cosmo_normalized_P}) on the cosmological constraints obtained by the analysis of the \planck\ tSZ power spectrum using the model defined by Eq. (\ref{eq:ClSZ_cosmo_analysis}).

\begin{figure*}
\centering
\includegraphics[height=4.4cm]{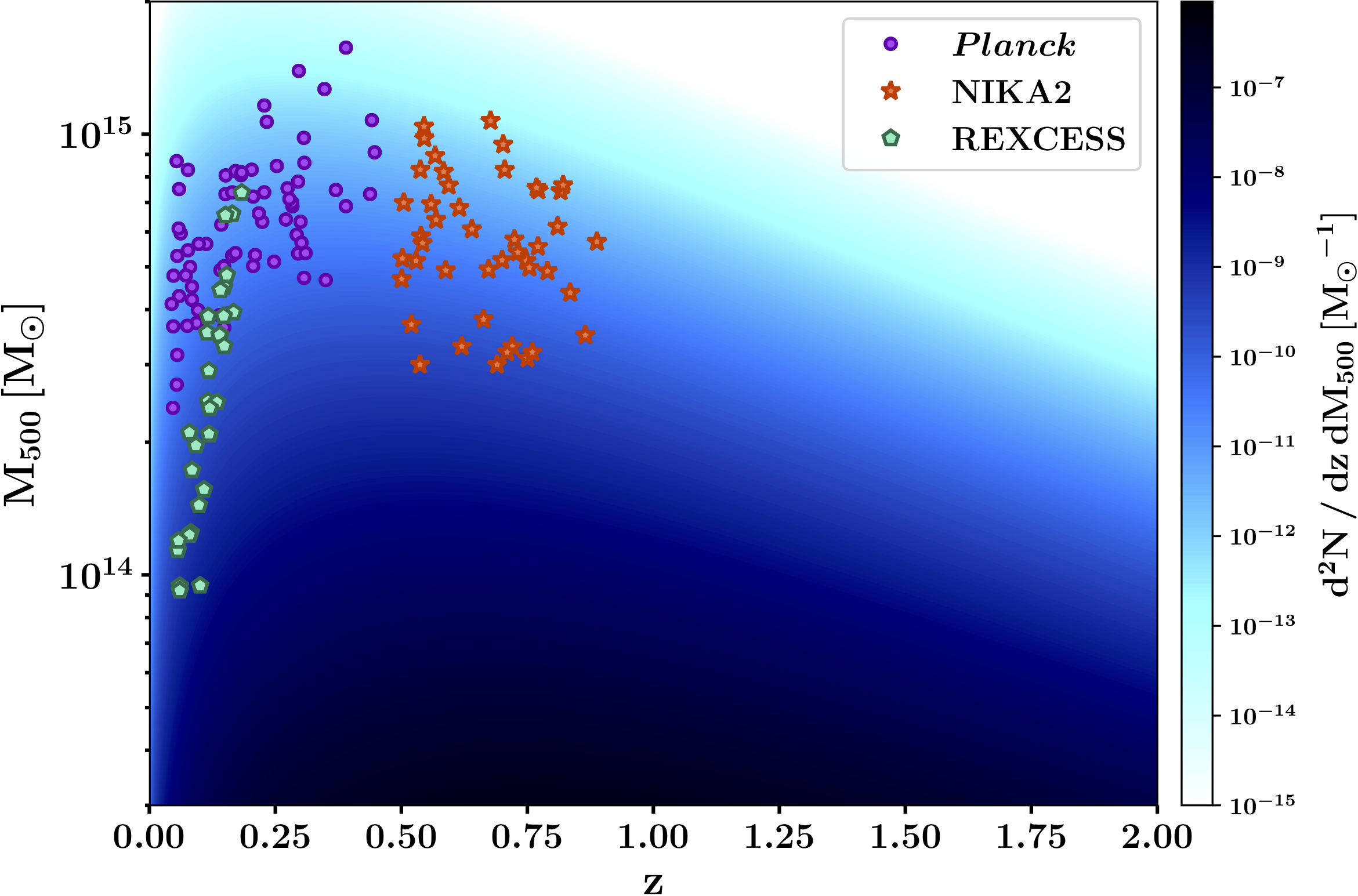}
\hspace{0.2cm}
\includegraphics[height=4.4cm]{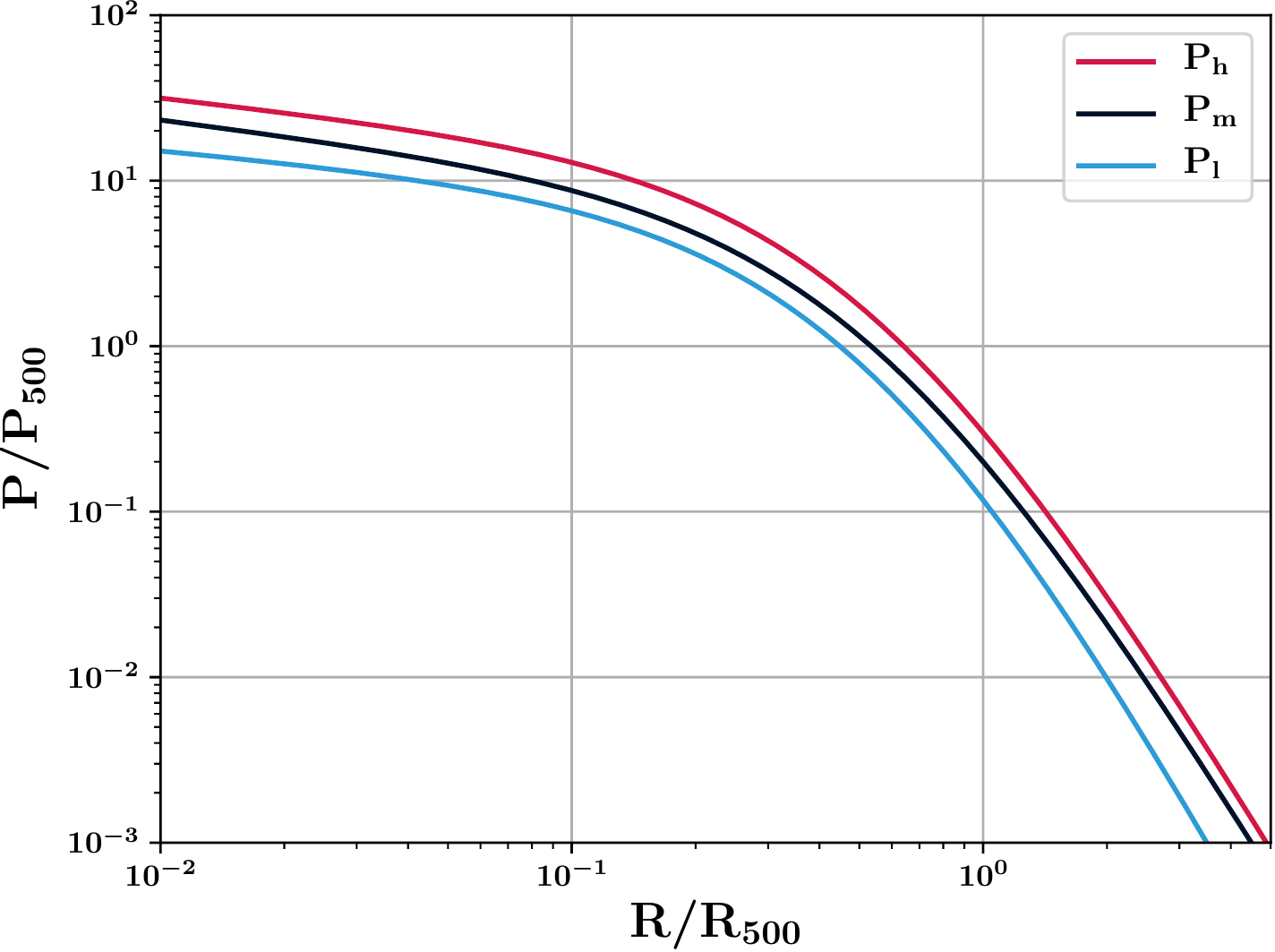}
\caption{{\footnotesize\textbf{Left:} Cluster abundance as a function of mass and redshift (shades of blue). We also show the distributions of the 62 \planck\ clusters (purple), the REXCESS (green), and NIKA2 (orange) samples. \textbf{Right:} Mean normalized pressure profiles considered in the analysis of the \planck\ tSZ power spectrum. The profile obtained by \planck\ is shown in black \citep{pla13}. The definition of the blue and red profiles is described in Sect. \ref{subsec:pressure_profile} \citep{rup19}.}}
\label{fig:context_P_prof}
\end{figure*}

\subsection{Selection of the mean normalized pressure profile}\label{subsec:pressure_profile}

The blue gradient in the left panel of Fig. \ref{fig:context_P_prof} shows the expected number of clusters per unit of mass and redshift assuming the cosmological parameters given in \cite{pla18}. The low redshift ($z < 0.5$) cluster samples REXCESS \citep{pra09} (green symbols) and \planck\ \citep{pla13} (purple dots) have been considered to establish the most widely used normalized pressure profiles in cosmological analyses. However, the dominant contribution to the tSZ power spectrum comes from the sum of the tSZ signal of low-mass halos at $z \gtrsim 0.3$ because their abundance is ${\sim}3$ times higher than the well-known massive clusters at low redshift \citep{pla16b}. If cluster self-similarity is not verified across the whole mass-redshift plane, the true mean normalized pressure profile of the cluster population will be different from the ones estimated at low redshift. The cluster sample of the NIKA2 SZ large program \citep{per18} (orange stars) has been defined in order to probe a potential redshift evolution of the mean pressure profile.\\
It is necessary to estimate the impact of a modification of this profile on the estimation of the $\sigma_8$ and $\Omega_m$ cosmological parameters. We show in the right panel of Fig. \ref{fig:context_P_prof} the two mean pressure profiles (blue and red lines) that we have defined on either side of the one obtained from the analysis of the 62 nearby clusters in the \planck\ sample (black line). We have used the gNFW parametric model given by Eq. (\ref{eq:cosmo_normalized_P}) to define these profiles. Their corresponding parameters have been set such that the profiles are compatible with the distribution of normalized pressure profiles obtained by \planck\ \citep{pla13} given its intrinsic scatter and that the corresponding gas mass fraction is compatible with the observed values at low redshift \cite{pla13,eck13,eck19}. We estimate the gas mass fraction profile for a given pressure profile by assuming that the cluster mass profiles are given by a Navarro-Frenk-White (NFW) model \citep{nav97}. We explain this procedure in detail in \citep{rup19}. The two pressure profiles defined with this procedure are called $P_l$ and $P_h$, whereas the \planck\ pressure profile is named $P_m$ in the following.

\begin{figure*}
\centering
\includegraphics[height=4.4cm]{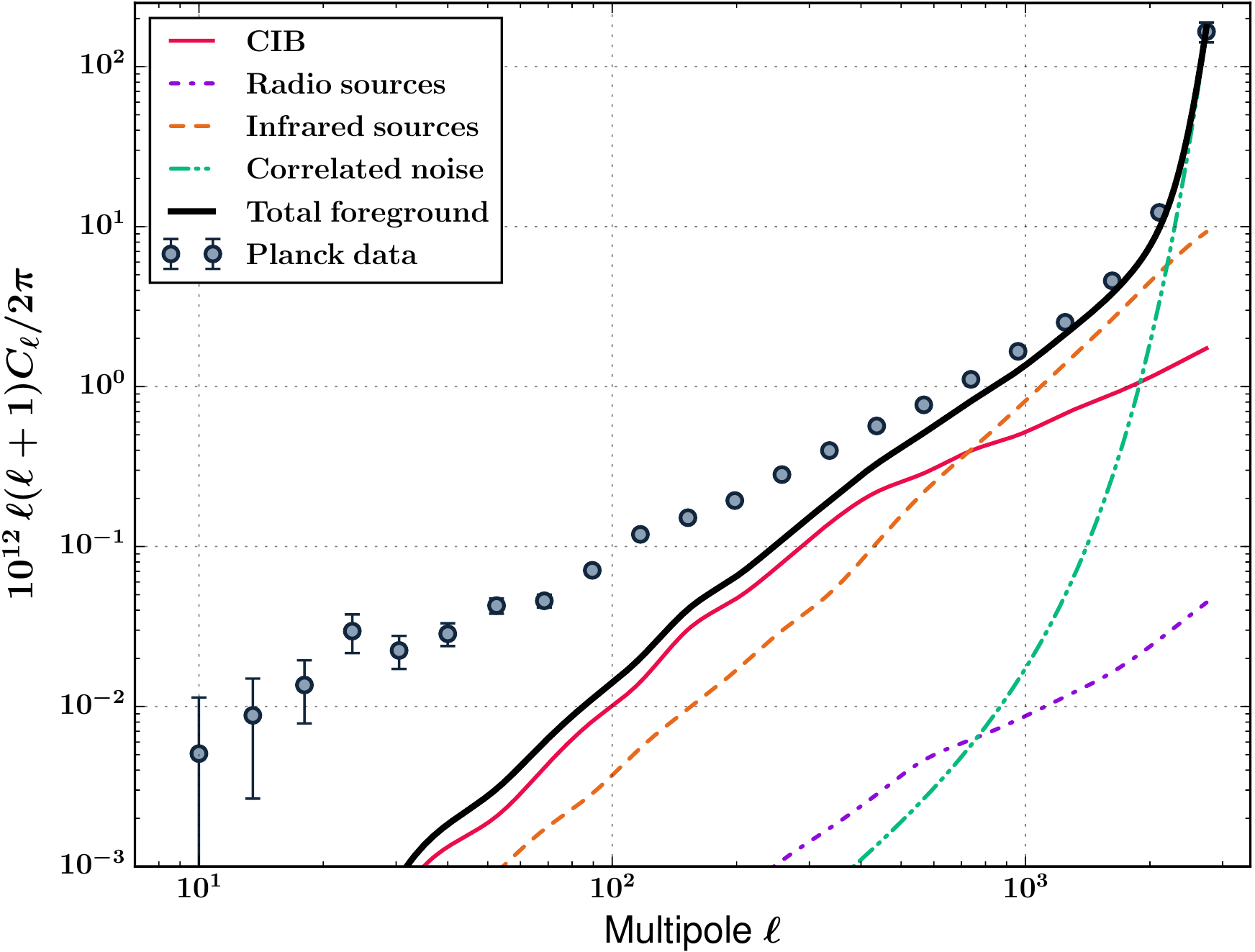}
\hspace{0.2cm}
\includegraphics[height=4.6cm]{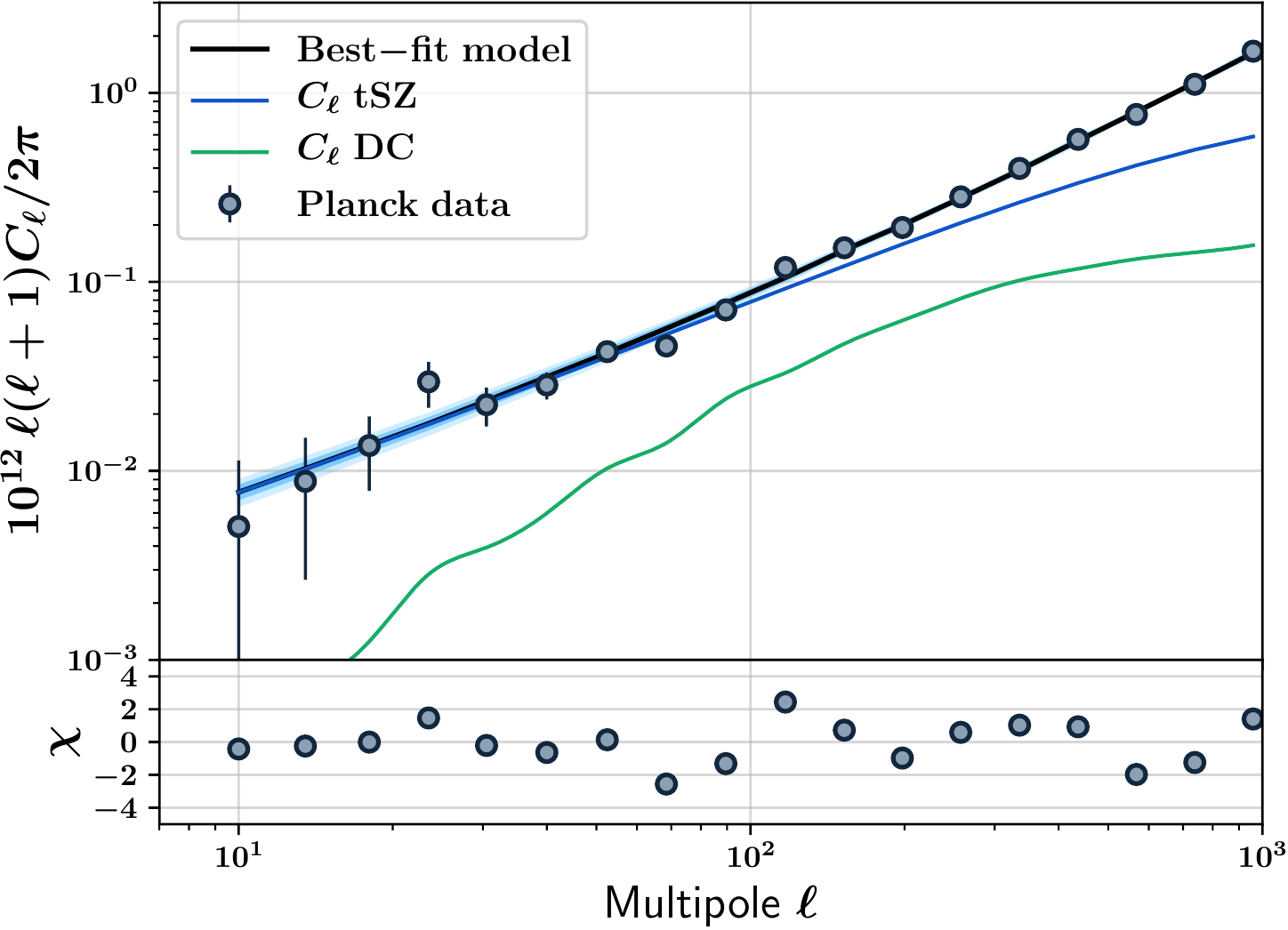}
\caption{{\footnotesize\textbf{Left:} \planck\ tSZ angular power spectrum measured from the sky $y$-map \citep{pla16b} (grey points). The power spectra of the contaminants to the tSZ signal are given by the colored lines and their sum is shown as a black line. \textbf{Right:} Best-fit model of the \planck\ tSZ power spectrum (black line). The blue line shows the contribution from the tSZ signal to the total power spectrum and the green line is the one due to the detected clusters in the \planck\ catalog  \citep{pla16}. The lower panel shows the significance of the residuals after the best-fit model subtraction to the data \citep{rup19}.}}
\label{fig:profile_bestfit}
\end{figure*}

\section{Analysis of the {\it\bf{{Planck}}} tSZ power spectrum}\label{sec:planck_tsz_power}

The power spectrum $C_{\ell}^{\mathrm{map}}$ of the \planck\ map of the tSZ effect \citep{pla16b} is represented by the grey points in the left panel of Fig. \ref{fig:profile_bestfit}. The error bars $\Delta C_{\ell}^{\mathrm{map}}$ associated with each bin in the power spectrum have been estimated using the cross-power spectra between the sky maps obtained by different detectors of the \planck\ satellite\footnote{We do not include the contribution from the trispectrum in the covariance matrix in this analysis although it has been shown to significantly increase the error bars of the power spectrum at low multipoles \citep{bol18}.}. All the contaminants to the tSZ signal are not completely excluded by the component separation analysis used to obtain the \planck\ map of the tSZ effect. The cosmic infrared background (CIB), unresolved infrared sources and radio sources induce significant residuals in the power spectrum. Spatially correlated instrumental noise also induces a significant increase in the power spectrum amplitude in the last multipole bins. Therefore, we model the total power spectrum by the following sum of components:
\begin{equation}
C_{\ell}^{\mathrm{map}} = C_{\ell}^{tSZ} + A_{\mathrm{CIB}}\hat{C}_{\ell}^{\mathrm{CIB}} + A_{\mathrm{IR}}\hat{C}_{\ell}^{\mathrm{IR}} + A_{\mathrm{RS}}\hat{C}_{\ell}^{\mathrm{RS}} + A_{\mathrm{CN}}\hat{C}_{\ell}^{\mathrm{CN}}
\label{eq:Cl_model_tot_cosmo}
\end{equation}
where the $\hat{C}_{\ell}^{i}$ are the power spectra of the $i^{th}$ component and $A_i$ their corresponding amplitudes. Tabulated models established by the \planck\ collaboration \citep{pla16b} for the power spectrum of the radio sources (RS), infrared sources (IR), CIB, and correlated noise (CN) have been used throughout this analysis. They are represented by colored lines in the left panel of Fig. \ref{fig:profile_bestfit} and their sum is shown as a black line. The dominant contribution to the power spectrum at multipoles $\ell \gtrsim 1000$ comes from these contaminants. Therefore, we choose to discard the bins with marginal information on the tSZ power spectrum and fit $C_{\ell}^{\mathrm{map}}$ up to the multipole bin $\ell = 959.5$ to optimize the computation time in our analysis.\\
\begin{figure*}
\centering
\includegraphics[height=4.9cm]{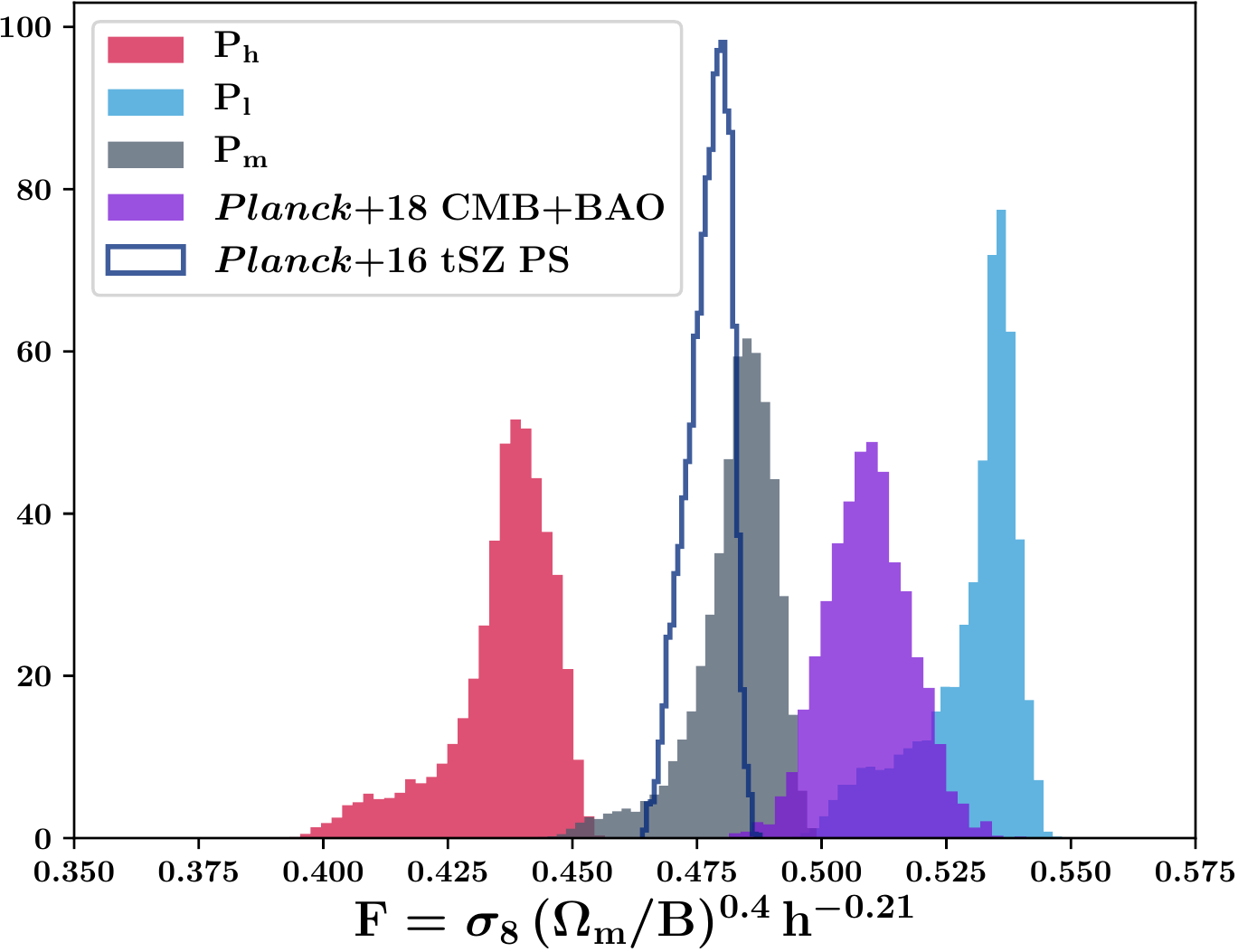}
\hspace{0.2cm}
\includegraphics[height=4.9cm]{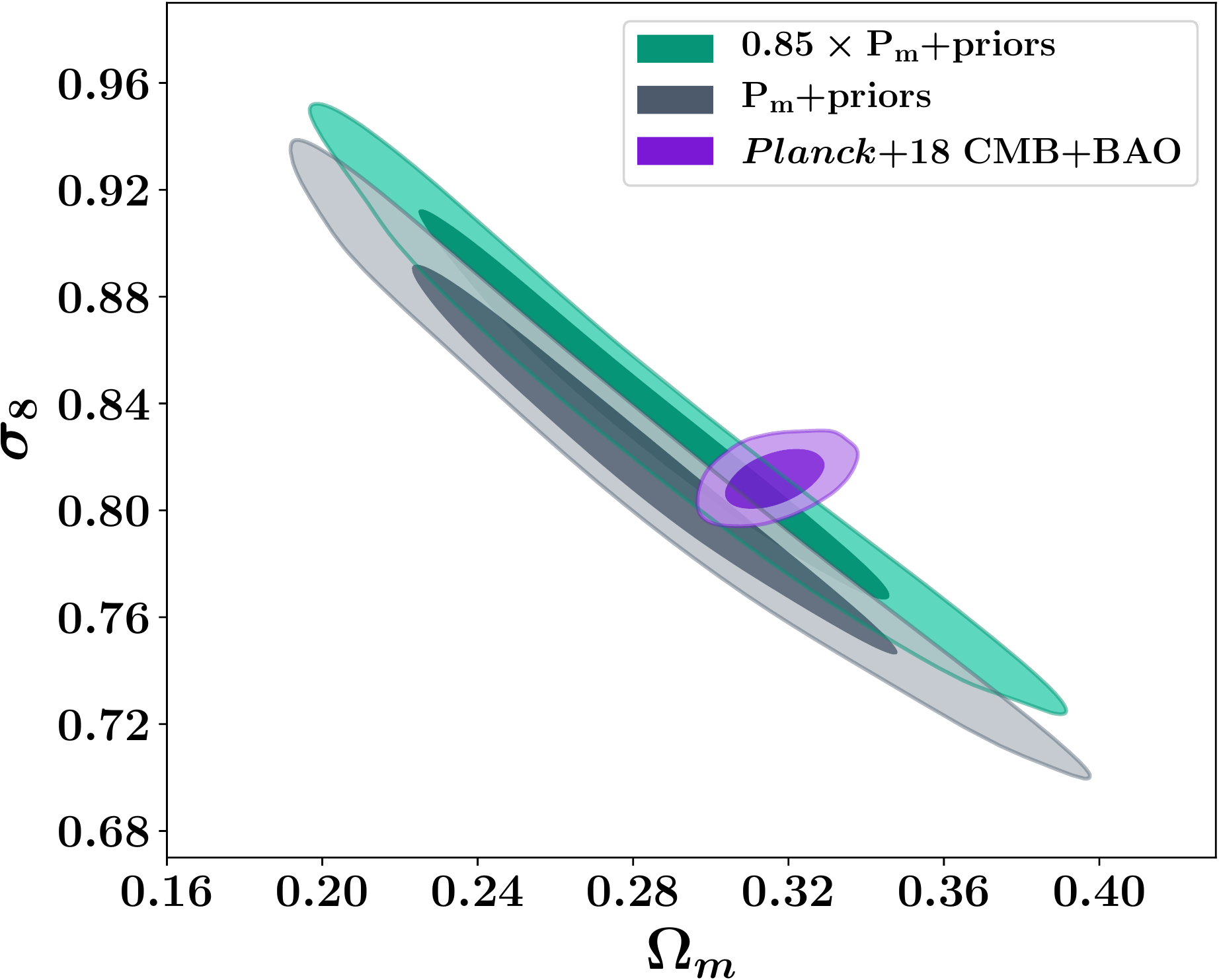}
\caption{{\footnotesize\textbf{Left:} Distributions of $F$ obtained at the end of the MCMC analyses based on the $P_h$ (red), $P_m$ (grey), and $P_l$ (blue) profiles. The distribution estimated by the \planck\ collaboration is represented by a dark blue line. The distribution obtained from the \planck\ analysis of the CMB primary anisotropies is also shown in purple. \textbf{Right:} Cosmological constraints obtained on $\sigma_8$ and $\Omega_m$  using the $P_m$ profile (grey) and the $P_m$ profile scaled down by 15\% (green) for an hydrostatic bias prior of $b = 0.20 \pm 0.01$. The \planck\ CMB contours are also shown (purple) \citep{rup19}.}}
\label{fig:cosmo_constraints}
\end{figure*}The three normalized pressure profiles defined in Sect. \ref{subsec:pressure_profile} are used in a MCMC analysis to fit the \planck\ power spectrum $ C_{\ell}^{\mathrm{map}}$ \citep{pla16b}. The free parameters in the analysis are $F$, $\Omega_m$, $b$, $h$, $A_{\mathrm{CIB}}$, $A_{\mathrm{IR}}$ and $A_{\mathrm{RS}}$. We fix the value of the amplitude of the power spectrum of the correlated noise to the one of $ C_{\ell}^{\mathrm{map}}$ measured in the bin at $\ell = 2742$. We use Eq. (\ref{eq:ClSZ_cosmo_analysis}) and (\ref{eq:Cl_model_tot_cosmo}) at each step of the analysis to compute the total power spectrum $\hat{C}_{\ell}^{\mathrm{tot}}$ given the considered cosmological model and pressure profile. This power spectrum is then compared with the \planck\ power spectrum in the 18 bins at $\ell < 1000$ through the Gaussian likelihood function $\mathcal{L}$ defined by:
\begin{equation}
-2 \mathrm{ln} \, \mathcal{L} = \sum_{\ell} \left[\frac{C_{\ell}^{\mathrm{map}} - \hat{C}_{\ell}^{\mathrm{tot}}}{\Delta C_{\ell}^{\mathrm{map}}}\right]^2
\end{equation}
We use uniform priors to ensure that all the parameters stay positive and that the hydrostatic bias value is such that $0 < b < 0.4$. Furthermore, we use a Gaussian prior on the Hubble parameter $h = 0.67\pm 0.03$ given the latest results obtained by the \planck\ CMB analysis \citep{pla18}. After the convergence of the chains and the burn-in cut-off, the remaining samples are used to estimate the marginalized probability densities associated with each parameter. The best-fit power spectrum model obtained at the end of the analysis based on the $P_m$ profile is represented by a black line in the right panel of Fig. \ref{fig:profile_bestfit}. As shown in the lower panel the statistical significance of the residuals computed after subtracting the best fit model to the measured power spectrum is smaller than $3\sigma$ at each multipole. The same statistical accuracy is reached at the end of the analyses based on the $P_l$ and $P_h$ profiles.

\section{Implications of a modification of the mean pressure profile}\label{sec:impact_pressure}

The posterior distributions of the $F$-parameter are shown in Fig. \ref{fig:cosmo_constraints} for the profiles $P_h$ (red), $P_m$ (grey) and $P_l$ (blue). We compare the distribution of $F$ obtained for the $P_m$ profile with the one estimated from the analysis of the tSZ power spectrum performed by the \planck\ collaboration \citep{pla16b} based on the same profile (dark blue line) and find that they are both fully compatible. The mean values of the distributions of the $F$-parameter associated with the $P_h$ and $P_l$ are significantly different from the one found using the $P_m$ profile. This shows that the mean normalized pressure profile considered in the cosmological analysis has a significant impact on the best-fit value of $F$. The distribution of $F$ computed with the parameter chains obtained from the analysis of the \planck\ CMB primary anisotropies is represented in purple in Fig. \ref{fig:cosmo_constraints}. Its mean value is enclosed between the ones obtained with the $P_m$ and $P_l$ profiles. This shows that the current discrepancy found between the cosmological constraints obtained from the analysis of the CMB primary anisotropies and the cluster abundance for a fixed hydrostatic bias may be canceled if the amplitude of the mean pressure profile of the cluster population is slightly lower than the one constrained at low redshift.\\
We have also realized the MCMC analysis described in Sect. \ref{sec:planck_tsz_power} by considering a possible future hydrostatic bias prior of $b = 0.20 \pm 0.01$ and a Gaussian prior on the matter density $\Omega_m = 0.2 \pm 0.08$. The results found on the $\sigma_8$ and $\Omega_m$ parameters are presented in the right panel of Fig. \ref{fig:cosmo_constraints} for the $P_m$ profile (grey) and for the same profile scaled down by 15\% (green). The best-fit values of these parameters obtained with the $P_m$ profile is found at 2-$\sigma$ from the ones estimated with the analysis of the \planck\ CMB primary anisotropies (purple). However, we find that a 15\% decrease of the amplitude of the \planck\ mean normalized pressure profile would reconcile the maximum likelihood values of $\sigma_8$ and $\Omega_m$ with the CMB results using a hydrostatic bias that is fully compatible with the current estimates. This result shows that it is essential to characterize the properties of the pressure profile of galaxy clusters in different regions of the mass-redshift plane in order to take into account the systematic effects caused by a deviations from self-similarity in SZ cosmological analyses.

\section{Conclusions}\label{sec:conclusions}

The analysis of the \planck\ tSZ power spectrum enabled us to show that both the current uncertainties on the hydrostatic bias parameter and the mean normalized pressure profile induce systematic effects that are much greater than the magnitude of the discrepancy observed between the estimates of $\sigma_8$ and $\Omega_m$ from low and high redshift cosmological probes. In particular, we have shown that a 15\% decrease of the amplitude of the \planck\ mean normalized pressure profile is sufficient to alleviate this tension if the hydrostatic bias is constrained at the percent level to a value of 0.2. Studying the potential mass and redshift evolution of the cluster thermodynamic properties is essential to handle accurately these systematic effects in cosmological analyses. In this context, the NIKA2 tSZ large program will provide valuable insights concerning the redshift evolution of the pressure profile and its impact on cosmology.

\section*{Acknowledgements}
\small{This work has been partially funded by the ANR under the contract ANR-15-CE31-0017. Support for this work was provided by NASA through SAO Award Number SV2-82023 issued by the CXC, which is operated by the Smithsonian Astrophysical Observatory for and on behalf of NASA under contract NAS8-03060. FR would like to thank M. Arnaud, J-B. Melin, and especially B. Bolliet for very useful and interesting discussions. We acknowledge funding from the ENIGMASS French LabEx.}

%

\begin{thebibliography}{}
{\small
%
%
\bibitem[{Voit (2005)}]{voi05}
M. Voit, Rev. Mod. Phys. \textbf{77}, 207-258 (2005)

\bibitem[{Bocquet \emph{et~al.}(2019)}]{boc18}
S. Bocquet \emph{et~al.}, \apj\ \textbf{878}, 55 (2019)

\bibitem[Planck Collaboration (2016)]{pla16b}
Planck Collaboration, \emph{et al.}, \aap\ \textbf{594}, A22 (2016)

\bibitem[Planck Collaboration (2016b)]{pla16c}
Planck Collaboration,\emph{et al.}, \aap\ \textbf{594}, A24 (2016b)

\bibitem[{Arnaud \emph{et~al.}(2010)}]{arn10}
M. Arnaud \emph{et~al.}, \aap\ \textbf{517}, A92 (2010)

\bibitem[{{Planck Collaboration} \emph{et~al.}(2013)}]{pla13}
{Planck Collaboration}, \emph{et~al.}, \aap\ \textbf{550}, A131 (2013)

\bibitem[{Adam \emph{et~al.}(2018)}]{ada18}
R. Adam \emph{et~al.}, \aap\ \textbf{609}, A115 (2018)

\bibitem[{{Perotto} \emph{et~al.}(2018)}]{per18}
L. {Perotto} \emph{et~al.}, proceeding paper of the Cosmology session of the 53rd Rencontres de Moriond conference, arXiv:1808.10817 (2018)

\bibitem[{{Komatsu} \& Kitayama (1999)}]{kom99}
E. {Komatsu} \& T. Kitayama, \apj\ \textbf{526}, L1 (1999)

\bibitem[{{Komatsu} \& Seljak (2002)}]{kom02}
E. {Komatsu} \& U. Seljak, \mnras\ \textbf{336}, 1256 (2002)

\bibitem[{{Nagai} \emph{et~al.}(2007)}]{nag07}
D. {Nagai}, A.~V. {Kravtsov}, \& A. {Vikhlinin}, \apj\ \textbf{668}, 1 (2007)

\bibitem[{Bolliet \emph{et~al.}(2018)}]{bol18}
B. Bolliet \emph{et~al.}, \mnras\ \textbf{477}, 4957-4967 (2018)

\bibitem[{{Planck Collaboration} \emph{et~al.}(2018)}]{pla18}
{Planck Collaboration}, \emph{et~al.}, arXiv:1807.06209 (2018)

\bibitem[Planck Collaboration (2016)]{pla16}
Planck Collaboration \emph{et al.}, \aap\ \textbf{594}, A27 (2016)

\bibitem[{Pratt \emph{et~al.}(2009)}]{pra09}
G. W. Pratt \emph{et~al.}, \aap\ \textbf{498}, 361-378 (2009)
  
  \bibitem[{Eckert \emph{et~al.}(2013)}]{eck13}
D. Eckert \emph{et~al.}, \aap\ \textbf{551}, A23 (2013)

\bibitem[{Eckert \emph{et~al.}(2019)}]{eck19}
D. Eckert \emph{et~al.}, \aap\ \textbf{621}, A40 (2019)
  
\bibitem[{{Navarro} \emph{et~al.}(1997)}]{nav97}
J. F. {Navarro}, C.~S. {Frenk}, \& S. D. M. {White }, \apj\ \textbf{490}, 493-508 (1997)
  
 \bibitem[{Ruppin \emph{et~al.}(2019)}]{rup19}
F. Ruppin \emph{et~al.}, \mnras\ \textbf{490}, 784-796 (2019)
}
\end{thebibliography}
%
%

\end{document}